\documentclass[aps,twocolumn,preprintnumbers,prd,showpacs,nofootinbib,floats,floatfix,endfloats]{revtex4}
\usepackage{amsmath}
\usepackage{graphicx}
\usepackage{float}
\usepackage{array}
\usepackage{multirow}
\usepackage{color}

\newlength{\picheight}
\newlength{\picwidth}
\newcommand{\importpic}[1]{\includegraphics[width = \picwidth, trim = 30 0 30 0, clip = true]{#1}}

\begin{document}

\preprint{UCI-HEP-TR-2013-19}

\title{Sensitivity of a Future High Energy $e^+ e^-$ Collider to $Z^\prime$ Bosons}
\author{David Kapukchyan\vspace*{0.1cm}}
\affiliation{Department of Physics and Astronomy,
University of California, Irvine, CA 92697, USA\vspace*{0.1cm}}

\author{Tim M.P. Tait}
\affiliation{Department of Physics and Astronomy,
University of California, Irvine, CA 92697, USA\vspace*{0.1cm}}

\date{\today}
\pacs{}

\begin{abstract}

We consider the capability of a future high energy $e^+ e^-$ collider
to search for the $Z^\prime$ bosons associated with a $U(1)^\prime$
gauge extension of the Standard Model.  Even for $Z^\prime$ masses
well above the center of mass energy of the collider, the
rates of
production of $e^+ e^-$, $\mu^+ \mu^-$, unflavored jets, and $b$-tagged
jets can reveal the presence of virtual $Z^\prime$ exchange.  We
consider collider
configurations with center-of-mass energy
200, 300, and 500~GeV, and show how various classes of $Z^\prime$
model-lines can be probed with $100~{\rm fb}^{-1}$ of integrated luminosity.
We find that for order one couplings, $Z^\prime$ masses on the order of several
TeV can be discovered, depending on the collider energy.  Our results suggest
that a 200 GeV collision energy is unlikely to discover any of the $Z^\prime$
models we consider that would not have already been visible in late stages of
LHC running, but could still help refine understanding of the couplings and
distinguish between different models.  For $e^+ e^-$ collision energies of 300 GeV,
parameter space beyond the reach of the LHC becomes accessible.

\end{abstract}

\maketitle

\section{Introduction}

The Standard Model (SM) explains all the data from experiments so far,
but many aspects of its underlying structure remain mysterious.  As an immediate
example, it is clear that local $SU(3) \times SU(2) \times U(1)$ gauge invariance plays
a huge role in its formulation, but it is not clear why Nature made this choice, 
as opposed to some other one.  Given our lack of understanding, an
obvious place to look for physics beyond the SM is by extending the gauge sector,
and the simplest extension is an extra $U(1)^\prime$ group factor.
In fact, such an addition to the SM is also motivated by
theoretical considerations.  The gauge structure of the SM and its fermonic matter
content both fit preternaturally into an $SU(5)$ grand unified theory (GUT).  Larger
groups such as $SO(10)$ and $E_6$, motivated by UV completions of gravity,
often contain remnant $U(1)^\prime$s at low energies \cite{Langacker:1980js,Hewett:1988xc}.

The experimental hallmark of a $U(1)^\prime$ extension is the existence of an
electrically neutral spin one boson -- a $Z^\prime$.  Provided the $Z^\prime$ has
couplings with SM quarks and leptons, it can be produced as a resonance at
high energy hadron colliders \cite{Leike:1998wr}.  Current searches for the reaction 
$q \bar{q} \rightarrow Z^\prime \rightarrow \ell^+ \ell^-$ (where $\ell = e$ or $\mu$) at the
LHC have provided powerful bounds on this kind of scenario, requiring the
$Z^\prime$ mass to be larger than about 2.5~TeV, if it has order one couplings
to the SM fermions \cite{ATLAS:2013jma,CMS:2013qca}.  
More weakly coupled
$Z^\prime$s remain viable even at somewhat smaller masses.
Ultimately,
a $Z^\prime$ with order one couplings to both quarks and leptons
is expected to be discovered at a high luminosity $\sqrt{s} = 14$~TeV
LHC run provided $M_{Z^\prime}$ is less than about 4-5 TeV \cite{Godfrey:2013eta}.
While a high luminosity LHC has excellent prospects to detect a $Z^\prime$,
the relatively limited number of viable channels would nonetheless leave many questions
about its nature unresolved.

Lepton colliders offer an alternative perspective on $Z^\prime$ searches.  A
near-future $e^+ e^-$ collider is unlikely to reach LHC energies, but is
expected to achieve an unprecedented precision that would allow it to be
sensitive to virtual $Z^\prime$s whose masses are far beyond the center-of-mass
energy.  In fact, bounds from LEP II remain interesting and relevant constraints on 
realistic $Z^\prime$ models even in the current LHC era \cite{LEP:2003aa,Carena:2004xs}.

In this article, we examine the discovery potential of a future $e^+ e^-$
collider operating at $\sim 500$~GeV collision energy to discover a $Z^\prime$.
Unlike earlier studies \cite{Rizzo:1996rx,Rizzo:1998vf,Rizzo:2003ug,Freitas:2004hq,Basso:2009hf,Blaising:2012tz,Baer:2013cma}
we focus on a set of models which are somewhat orthogonal generalizations of the
usual $E_6$-inspired $Z^\prime$ models \cite{Carena:2004xs}, and focus on lower
energy running options, $\sqrt{s} = 200, 300,$ and $500$ GeV.  These energies,
while low enough to perhaps exclude the possibility of studying exotic states directly,
would nonetheless allow for precision studies of the newly discovered \cite{Aad:2012tfa,Chatrchyan:2012ufa}
Higgs boson \cite{Baer:2013cma,Peskin:2012we,Klute:2013cx}
and electroweak interactions of the top quark \cite{Baer:2013cma,Batra:2006iq}.

\section{$Z^\prime$ Models and Signal}

A completely generic $U(1)^\prime$ model is described by a large number of parameters, including the mass,
coupling, and width\footnote{The width is an important parameter in searches for on-shell $Z^\prime$s, such as
at the LHC, but is largely irrelevant for off-shell searches such as are more typical at $e^+ e^-$ colliders.}
of the $Z^\prime$, as well as the couplings to the SM fermions and Higgs boson.  We specialize to the
case in which the $Z^\prime$ charges of the fermions are generation-independent, motivated by the desire
to limit contributions to flavor-changing neutral currents at tree level \cite{Carena:2004xs}.  
The important couplings are thus those to the left-handed quark doublet ($g_{qL}$), the right-handed
up-type quarks ($g_{uR}$) and down-type quarks ($g_{dR}$), and to the lepton doublet ($g_{lL}$)
and charged singlet ($g_{lR}$).
We further avoid
strong electroweak precision constraints by choosing models for which the SM Higgs is uncharged, preventing
$Z$-$Z^\prime$ mixing at tree level.
Four interesting model-lines belonging to this class, and for which gauge anomalies can be cancelled by vector-like
(under the SM) fermions are defined in Ref.~\cite{Carena:2004xs}: 
$U(1)_{\rm universal}$, $U(1)_{B-xL}$, $U(1)_{10+x\bar{5}}$, and $U(1)_{d-xu}$.   Each of these model-lines  
is further characterized by a continuous real parameter $x$ which parameterizes a freedom to choose
the charges of the SM fermions under $U(1)^\prime$ while maintaing the properties outlined above.
The $E_6$ model points are contained as special values of $x$ (for some of the model-lines)
within this wider framework \cite{Carena:2004xs}.
In addition, the gauge coupling, which multiplies the charge in each interaction, and the $Z^\prime$ mass
are independent parameters.
The charges of the models in terms of $x$ are summarized in 
Table~\ref{table:model_summary}.

\begin{table}
 \begin{center}
  \begin{tabular}{| c | c | c | c | c |}
   \hline
	$Z^\prime$ Model & $U(1)_{\rm universal}$ & $U(1)_{B-xL}$ & $U(1)_{10+x\bar{5}}$ & $U(1)_{d-xu}$ \\
	\hline
	$g_{qL}$ & $x$ & $\frac{1}{3}$ & $\frac{1}{3}$ & $0$ \\
	\hline
	$g_{uR}$ & $x$ & $\frac{1}{3}$ & $-\frac{1}{3}$ & $-\frac{x}{3}$ \\
	\hline
	$g_{dR}$ & $x$ & $\frac{1}{3}$ & $-\frac{x}{3}$ & $\frac{1}{3}$ \\
	\hline
	$g_{lL}$ & $x$ & $x$ & $\frac{x}{3}$ & $\frac{-1+x}{3}$ \\
	\hline
	$g_{lR}$ & $x$ & $x$ & $-\frac{1}{3}$ & $\frac{x}{3}$ \\
   \hline
  \end{tabular}
  \caption{Summary of the fermion charges for the four $Z^\prime$ models considered, in terms
  of the continuous real parameter $x$.
  \label{table:model_summary}}
 \end{center}
\end{table}
\printtables

The LHC limits suggest that a $Z^\prime$ strongly enough coupled to be discoverable at a future
high energy $e^+ e^-$ facility must have a mass well above the collider energy.  As a result, it
contributes to scattering virtually, effectively acting as a higher dimensional interaction which interferes
with the SM contributions mediated by the photon and $Z$ boson.  We examine the rate for
$e^+ e^-$ annihilation into electrons, muons, jets containing bottom quarks, 
and unflavored jets of hadrons (initiated by $u$, $d$, $s$, or $c$ quarks).  We do not explicitly consider
$\tau$ production, which is sensitive to poorly understood $\tau$-tagging rates at future detectors,
or top production, which requires a more complicated analysis to identify the top decay products.
Both of these are worthwhile, but not likely to make up the bulk of a $Z^\prime$ signal
for our benchmark model-lines.  

The influence of the $Z^\prime$ on the rate of a given channel
depends on how it interferes with the SM amplitudes, leading either to an increase or depression
compared to the SM expectation.  We simulate the leading order rate with Madgraph~5 \cite{Alwall:2011uj},
with the showering and hadronization performed by Pythia \cite{Sjostrand:2007gs}
and then simulate the detector response using PGS \cite{PGS}.  

We examine three different center-of-mass energies for the $e^+ e^-$ collisions,
$\sqrt{s} = 200, 300,$ and 500~GeV and a total data set of ${\cal L} = 100~{\rm fb}^{-1}$
integrated luminosity.  For each energy, we determine the expected number of events
in each channel assuming the SM is correct, and that the uncertainty in this quantity
is dominated by the statistical error\footnote{This was the case for the LEP II $Z^\prime$
analyses \cite{LEP:2003aa}.}.  We consider a given parameter point discoverable if it leads to
a deviation of $5\sigma$ in the expected rate.  We draw contours in the plane of the
$Z^\prime$ mass versus $x$ for fixed $U(1)^\prime$ gauge coupling $g=1$, delineating the regions
of parameter space in which a $Z^\prime$ is observable.
It is worth noting that for much of the parameter space considered here, a high luminosity LHC
would presumably discover the $Z^\prime$ before a future high energy $e^+ e^-$ collider begins operation.
We choose $5\sigma$ as a case where the $e^+ e^-$ collider can definitively identify the presence of
the $Z^\prime$ in a given channel and also measure the deviation in that channel to order $20\%$.

\section{Results}

\begin{figure}
\importpic{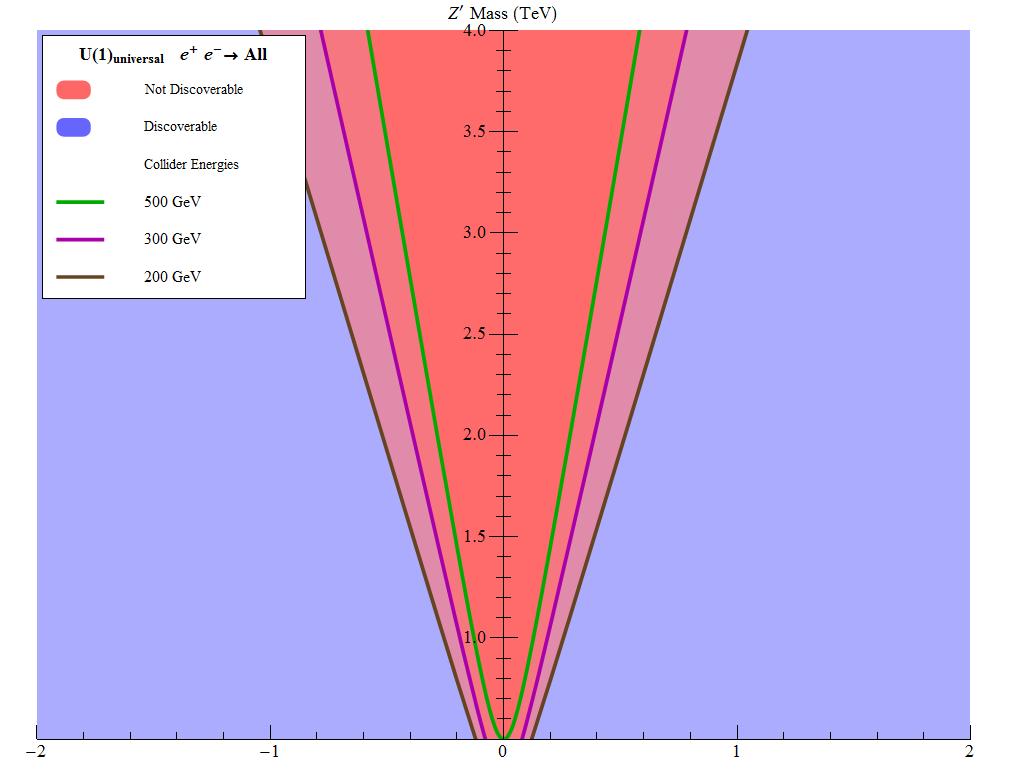}
\caption{The discoverable region of $U(1)_{\rm universal}$ for an $e^+ e^-$ collider
running at 200, 300, or 500 GeV (as indicated) having collected 100 ${\rm fb}^{-1}$
of integrated luminosity, in the plane of the coupling and the $Z^\prime$ mass, based on
the inclusive rate into SM fermions.
\label{fig:model_equality_TCx}
}
\end{figure}

\begin{figure}
\importpic{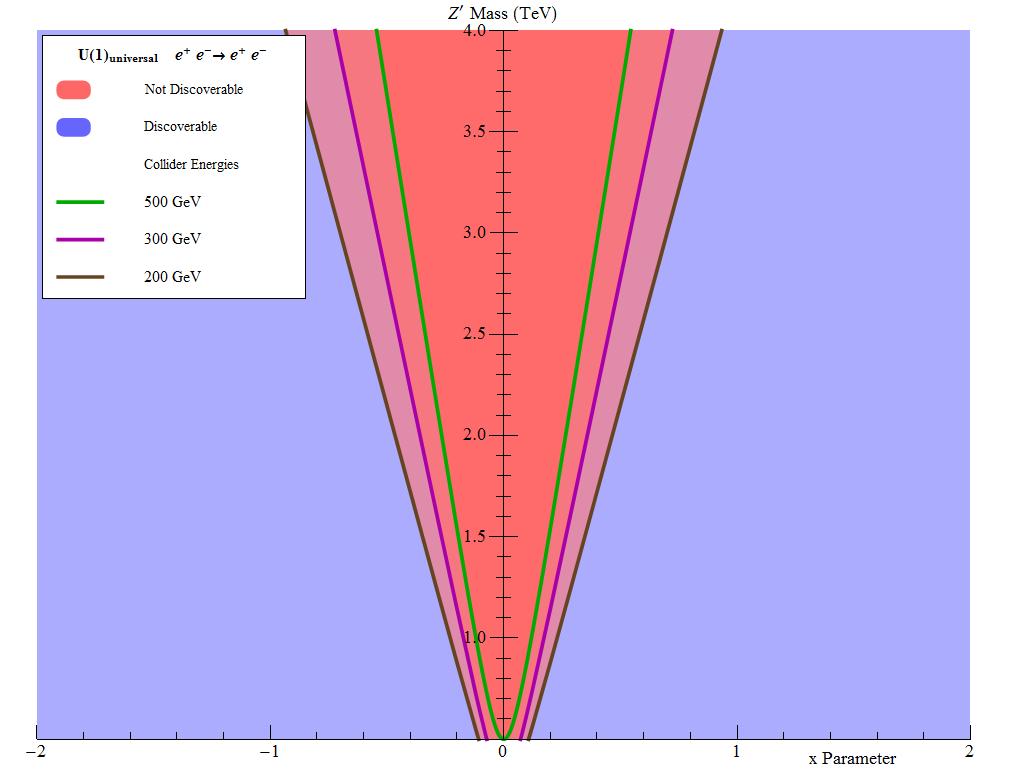}
\caption{Same as Figure \ref{fig:model_equality_TCx}, except for the exclusive $e^+ e^-$ final state.
\label{fig:model_equality_eCx}
}
\end{figure}

\begin{figure}
\importpic{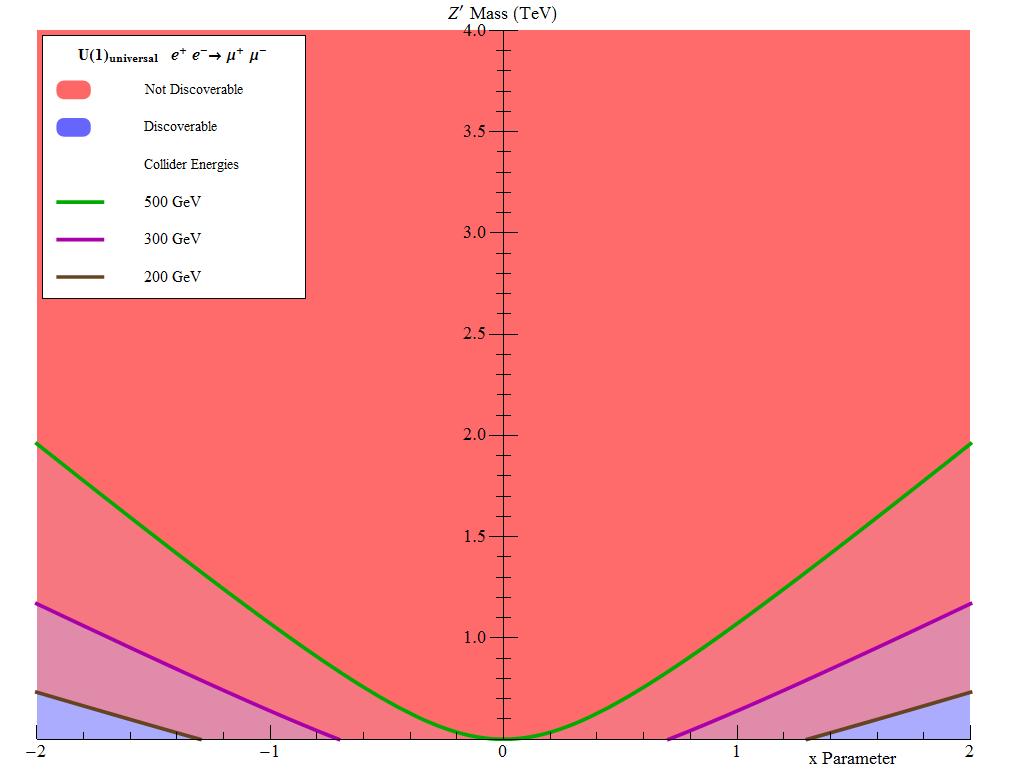}
\caption{Same as Figure \ref{fig:model_equality_TCx}, except for the exclusive $\mu^+ \mu^-$ final state.
\label{fig:model_equality_muonCx}
}
\end{figure}

\begin{figure}
\importpic{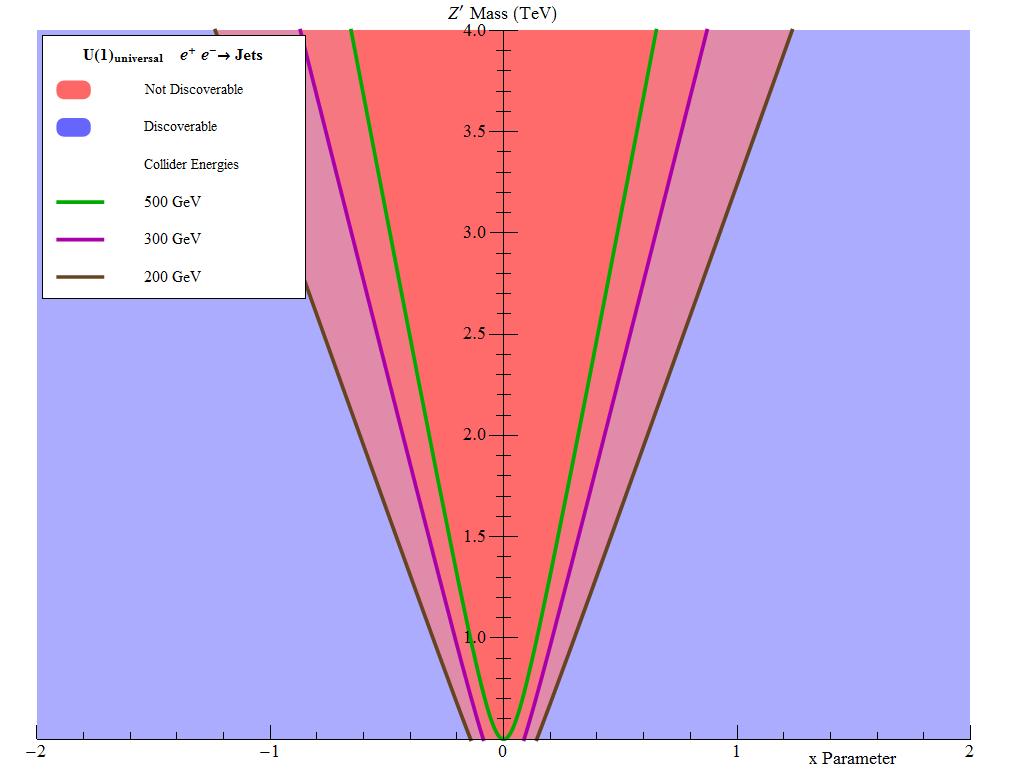}
\caption{Same as Figure \ref{fig:model_equality_TCx}, except for the exclusive unflavored jet final state.
\label{fig:model_equality_JCx}
}
\end{figure}

\begin{figure}
\importpic{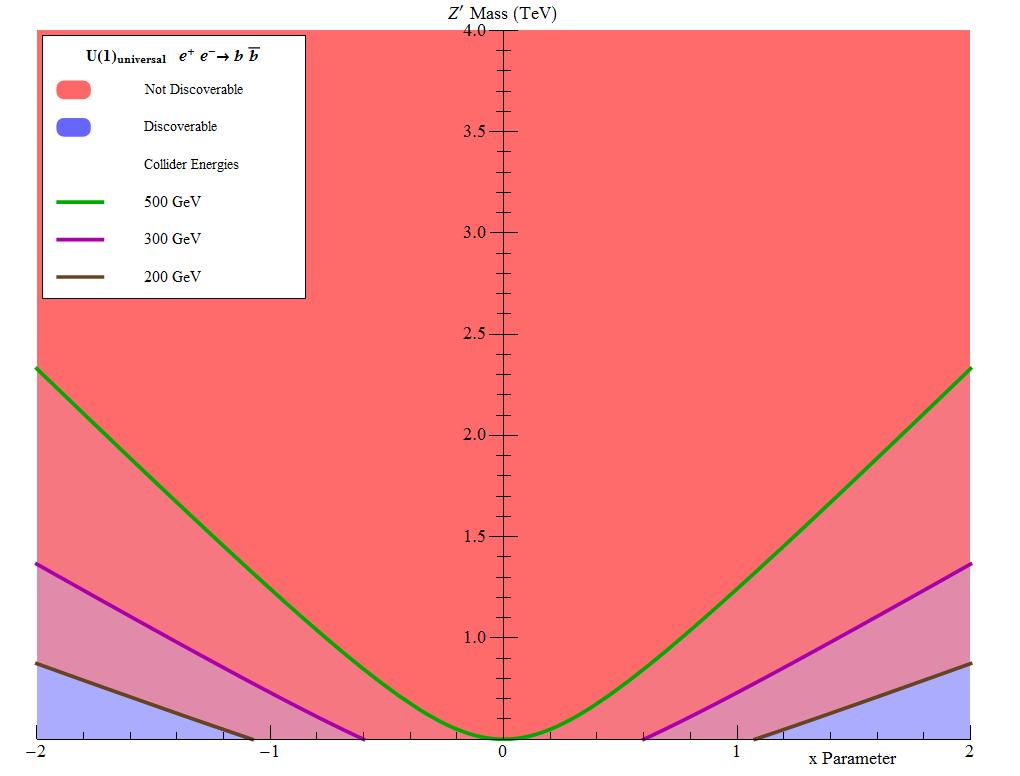}
\caption{Same as Figure \ref{fig:model_equality_TCx}, except for the exclusive $b$-jet final state.
\label{fig:model_equality_BCx}
}
\end{figure}

We summarize our findings in a series of Figures for each $Z^\prime$ model, showing the bounds that can be
placed on each model-line in the parameter space of the $Z^\prime$ mass and $x$, for each production
channel individually, as well as the inclusive rate of all channels combined.
In Figures~\ref{fig:model_equality_TCx}-\ref{fig:model_equality_BCx}, we show the bounds
on $U(1)_{\rm universal}$, where all fermions charges are equal to $x$ (which may then itself be interpreted as the
$U(1)$ gauge coupling).   The plots illustrate regions where a $5\sigma$ deviation from the SM expected rate
is realized,
in the plane of the $Z^\prime$ mass and the coupling (equivalent to $x$ for $U(1)_{\rm universal}$).  
We choose not
to present results for $Z^\prime$ masses below 500 GeV, where on-shell production becomes possible, and the bounds
depend sensitively on the $Z^\prime$ width.

For $U(1)_{\rm universal}$, the strongest bounds at a given collider energy
come from the electron channel, largely because of the fact that the
$Z^\prime$ can contribute via both $t$-channel as well as $s$-channel Feynman diagrams.  As a result, the inclusive
and electron-specific bounds are similar.  The results indicate that even at 200 GeV, $Z^\prime$ masses
up to around 4 TeV can be discovered for order one couplings, with the difficult-to-discover region lying in the regime of
small coupling and large mass.  This parameter space is likely to have been already covered by the high energy LHC
run \cite{Godfrey:2013eta,Osland:2009dp}, but already at 300 GeV, the $e^+ e^-$ collider is expected to have a somewhat longer reach.

\begin{figure}
\importpic{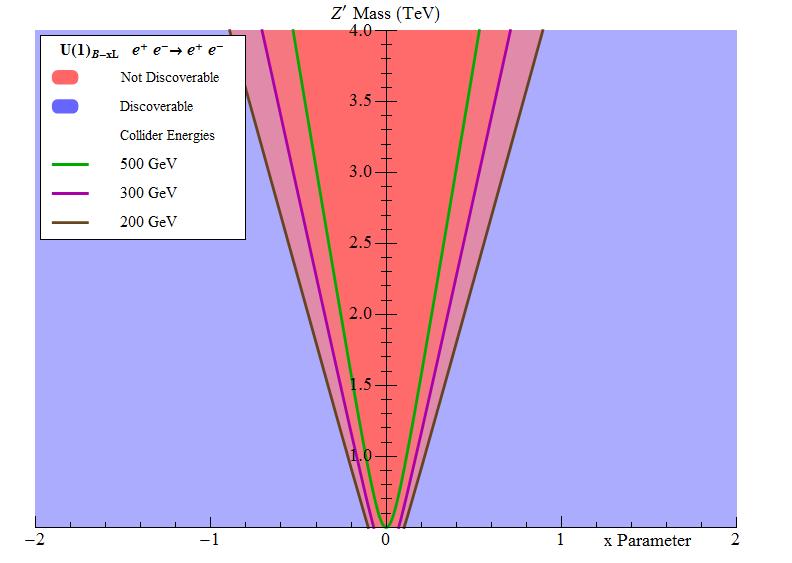}
\caption{The discoverable region of $U(1)_{B-xL}$ for an $e^+ e^-$ collider
running at 200, 300, or 500 GeV (as indicated) having collected 100 ${\rm fb}^{-1}$
of integrated luminosity, in the plane of the model parameter $x$ and the $Z^\prime$ mass 
(for fixed gauge coupling $g=1$), 
based on the $e^+ e^-$ final state.
\label{fig:model_inequality_eCx}
}
\end{figure}

\begin{figure}
\importpic{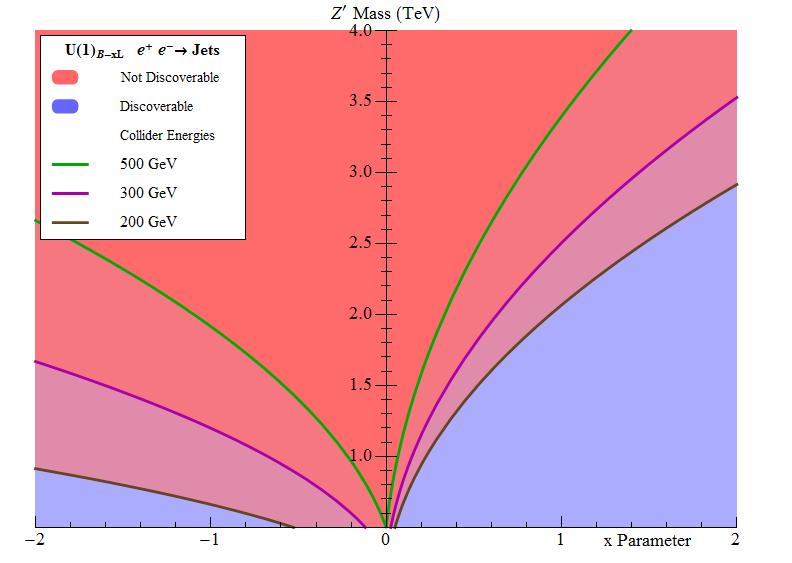}
\caption{Same as Figure~\ref{fig:model_inequality_eCx}, except for the exclusive unflavored jet final state.
\label{fig:model_inequality_JCx}
}
\end{figure}

\begin{figure}
\importpic{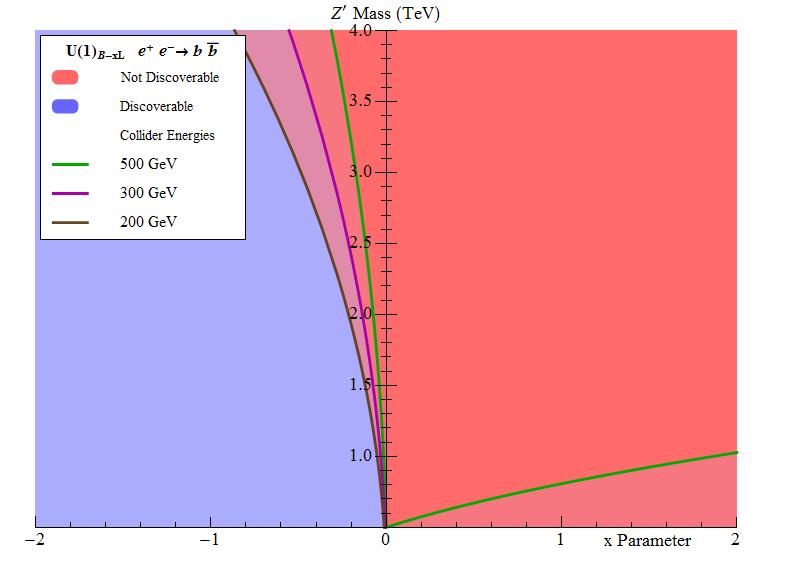}
\caption{Same as Figure~\ref{fig:model_inequality_eCx}, except for the exclusive $b$-jet final state.	\label{fig:model_inequality_BCx}
}
\end{figure}

These trends continue to be evident in other types of $Z^\prime$ models.  $U(1)_{B-xL}$
is very similar to the model with universal couplings, except for $x \rightarrow 0$ (where
searches involving leptons in the initial or final states are hopeless).  As shown in 
Figure~\ref{fig:model_inequality_eCx}, the coverage based on the electron final state is
essentially the same as for $U(1)_{\rm universal}$, and this limit continues to dominate the
other channels for the same reasons.  One different feature is the fact that the over-all
sign of the $Z^\prime$ contribution to jet final states
is controlled by the sign of the parameter $x$, leading to very different results
(shown in Figures~\ref{fig:model_inequality_JCx} and \ref{fig:model_inequality_BCx})
for positive and negative $x$ values.  The differences in rates would be helpful in
disentangling the nature of a $Z^\prime$ signal.  Even in cases where one would expect the LHC
to have made an initial discovery and measured $|x|$, the sign of $x$ can be determined
even with relatively modest $e^+ e^-$ collision energy.

\begin{figure}
\importpic{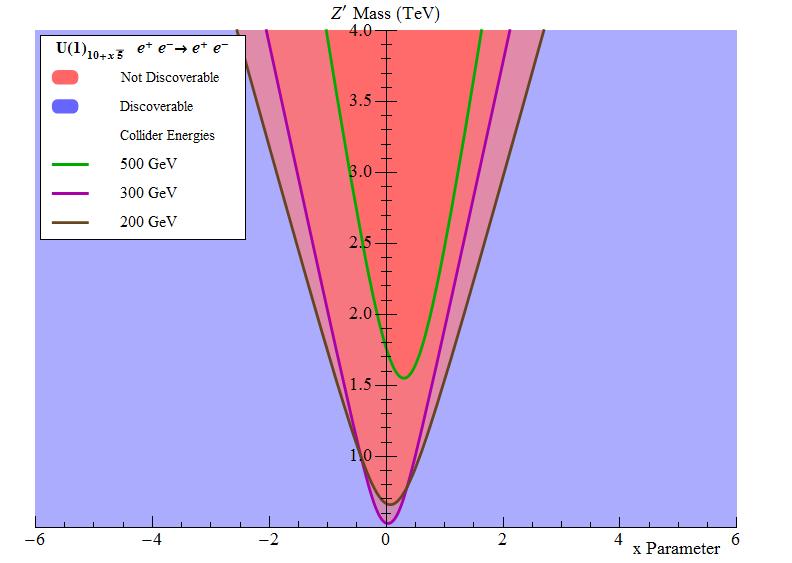}
\caption{The discoverable region of $U(1)_{10+x\bar{5}}$ for an $e^+ e^-$ collider
running at 200, 300, or 500 GeV (as indicated) having collected 100 ${\rm fb}^{-1}$
of integrated luminosity, in the plane of the model parameter $x$ and the $Z^\prime$ mass
(for fixed gauge coupling $g=1$), 
based on the $e^+ e^-$ final state.
\label{fig:model_3_eCx}
}
\end{figure}

\begin{figure}
\importpic{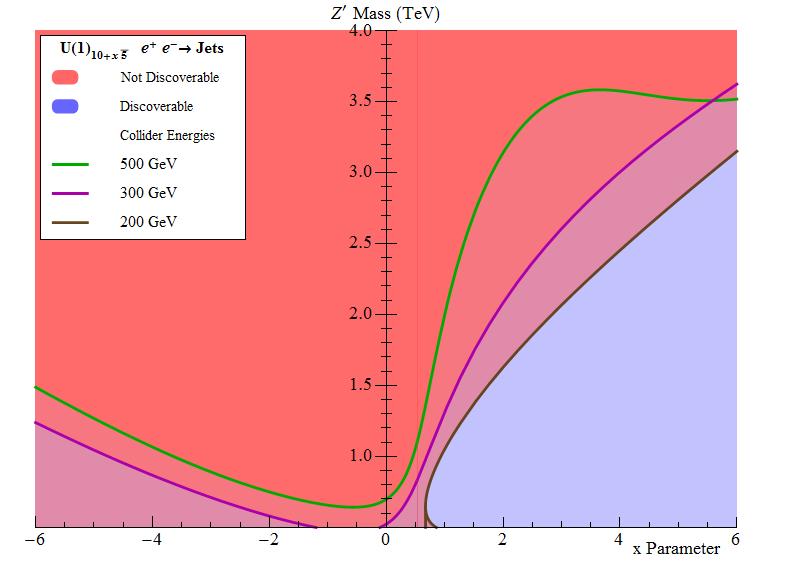}
\caption{Same as Figure~\ref{fig:model_3_eCx}, except for the exclusive unflavored jet final state.
\label{fig:model_3_JCx}
}
\end{figure}

\begin{figure}
\importpic{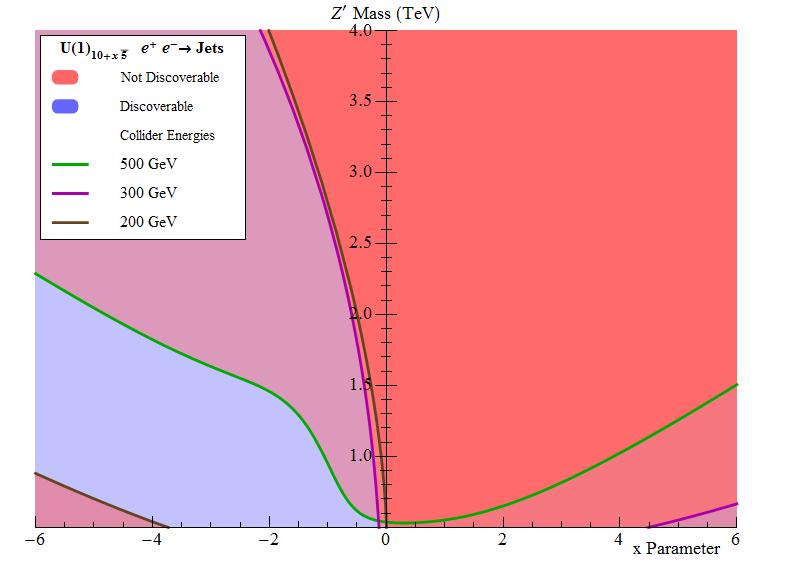}
\caption{Same as Figure~\ref{fig:model_3_eCx}, except for the exclusive $b$-jet final state.	\label{fig:model_3_BCx}
}
\end{figure}

The results for $U(1)_{10+x\bar{5}}$ illustrate a case where the couplings vary more widely with the parameter
$x$.  The $e^+ e^-$ final state, illustrated in Figure~\ref{fig:model_3_eCx}, continues to provide the over-all best
limit for most values of $x$.  The fact that the coupling to right-handed leptons is fixed, whereas the sign of the
coupling to $g_{lL}$ is determined by the sign of $x$; introduces a noticeable asymmetry to the discoverable region,
which becomes much more pronounced in the reach using untagged jets (Figure~\ref{fig:model_3_JCx}),
and $b$-tagged jets (Figure~\ref{fig:model_3_BCx}).  Once again, it is clear that valuable information about
the pattern of couplings to various SM fermions can be inferred from a multi-channel analysis, giving clues as to
the nature of a $Z^\prime$ signal if one is observed.  This model also illustrates the utility of running at different energies,
which changes the interplay between the $Z^\prime$ exchange amplitude squared compared with its interference with
the Standard Model amplitudes.  At low $Z^\prime$ masses, the two are comparable and regions of parameter space
exist where, for example (in the $e^+ e^-$ final state),
the collider configuration with $\sqrt{s} = 200$~GeV can have a slightly improved reach compared
to the $300$~GeV configuration, owing to cancellations between the two terms.

\begin{figure}
\importpic{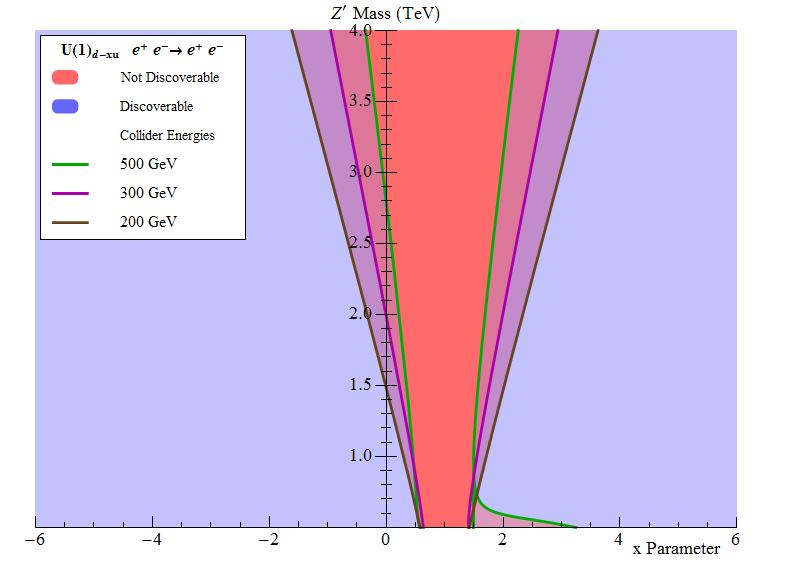}
\caption{The discoverable region of $U(1)_{d-xu}$ for an $e^+ e^-$ collider
running at 200, 300, or 500 GeV (as indicated) having collected 100 ${\rm fb}^{-1}$
of integrated luminosity, in the plane of the model parameter $x$ and the $Z^\prime$ mass
(for fixed gauge coupling $g=1$), 
based on the $e^+ e^-$ final state.
\label{fig:model_4_eCx}
}
\end{figure}

Similar features are present in $U(1)_{d-xu}$, shown in Figures~\ref{fig:model_4_eCx}.  
The $e^+ e^-$ final state remains the single most effective channel, and there are regions
of parameter space around $x \simeq 2$ and $Z^\prime$
masses slightly above 500 GeV where the destructive interference at the 300 or 500 
GeV collider configuration weakens discovery potential compared to the 200 GeV configuration.

%Figure command begins here%

%Figure command ends here

\section{Interplay with LHC}

As noted above, even in cases where the LHC would be likely to discover a $Z^\prime$ before a future $e^+ e^-$
collider begins operation, the $e^+ e^-$ environment can contribute much to refine the understanding
of the $Z^\prime$ model.

As an example, we consider the case of $U(1)_{B-xL}$ with $x=1$ and $M_{Z^\prime} = 3.5$~TeV.
Such a $Z^\prime$ would not have been discovered in the current LHC data set, but is likely to be
visible through the reaction $p p \rightarrow \ell^+ \ell^-$ in high
luminosity running at $\sqrt{s} = 14$~TeV \cite{Godfrey:2013eta}.  This observation would
unambiguously establish the existence of a new resonance coupled to both quarks and leptons.
However, even with high luminosity, the LHC is unlikely to be able to sift the decay into unflavored
dijets from the large backgrounds \cite{Yu:2013wta}, and for these parameters
observation of a decay into $t \bar{t}$ is expected to be marginal \cite{Barger:2006hm,Agashe:2007ki}.
The LHC observation would provide a measurement of the $Z^\prime$ mass, but its width would be
far below the experimental resolution, and thus only weakly bounded from above.

Even at $\sqrt{s} = 200$~GeV and with 100 ${\rm fb}^{-1}$, a future $e^+ e^-$ collider would do much
to clarify the nature of such an LHC discovery.  From Figures \ref{fig:model_inequality_eCx}-\ref{fig:model_inequality_BCx},
we see that the expectation is that such a machine would observe only the $e^+ e^- \rightarrow e^+ e^-$ channel,
narrowing down the space of possible explanations to those where the coupling to leptons is roughly comparable
to the coupling to quarks.  Comparison with Figure~\ref{fig:model_equality_JCx} indicates that the model
with universal couplings would be strongly disfavored by the lack of a signal in jets.  In terms of
an interpretation within the framework of 
$U(1)_{B-xL}$, such an observation would provide a measurement of $0.9 \leq |x| \leq 1.1$.  In fact,
because the $e^+ e^-$ collider's primary observable is through interference with the SM process,
the rate also provides information about the relative signs of the couplings.  For example, from
Figure~\ref{fig:model_inequality_BCx}, we see that for $x \simeq 1$, no signal in $b$-flavored jets is
expected to be visible, whereas for $x \simeq -1$, a signal should be strongly visible.

For this particular example, even a low energy $e^+ e^-$ collider is able to essentially reconstruct much
of the structure of the $Z^\prime$ couplings through simple rate information.  If polarized beams or
forward-backward asymmetries are employed, one can further dissect the chiral structure of
the couplings \cite{Osland:2009dp,Petriello:2008zr,Li:2009xh} (which in this example are
vector-like).

\section{Conclusions}

We have examined the discovery potential of a future $e^+ e^-$ collider to infer the existence of a $Z^\prime$
based on its virtual influence over the rate of SM fermions production,
looking at four well-motivated $Z^\prime$ model-lines.  We find that even at very modest
energies of $\sqrt{s} \simeq 300$~GeV, the discovery potential is somewhat higher than for
the LHC at design energy, and valuable information about the nature of the $Z^\prime$ couplings can
be inferred even for models that the LHC is expected to have already discovered.
We have confined our discussion to the simplest rate-based observables; angular variables
and polarized beams can extract even more information \cite{Osland:2009dp,Petriello:2008zr,Li:2009xh}.
We leave such refinements for future work.

A future $e^+ e^-$ facility has the potential to transform our understanding of the Higgs boson.  In this work, we see
that it also has interesting potential to search for physics beyond the Standard Model.

\section*{Acknowledgments}
The research of T.M.P.T. is supported in part by NSF
grants PHY-0970171 and PHY-1316792 and by the University of California, Irvine through a Chancellor's fellowship.


\newpage

\begin{thebibliography}{1}

  %\cite{Langacker:1980js}
\bibitem{Langacker:1980js} 
  P.~Langacker,
  %``Grand Unified Theories and Proton Decay,''
  Phys.\ Rept.\  {\bf 72}, 185 (1981).
  %%CITATION = PRPLC,72,185;%%
  %1118 citations counted in INSPIRE as of 29 Oct 2013

%\cite{Hewett:1988xc}
\bibitem{Hewett:1988xc} 
  J.~L.~Hewett and T.~G.~Rizzo,
  %``Low-Energy Phenomenology of Superstring Inspired E(6) Models,''
  Phys.\ Rept.\  {\bf 183}, 193 (1989).
  %%CITATION = PRPLC,183,193;%%
  %996 citations counted in INSPIRE as of 29 Oct 2013
  
  %\cite{Leike:1998wr}
\bibitem{Leike:1998wr} 
  A.~Leike,
  %``The Phenomenology of extra neutral gauge bosons,''
  Phys.\ Rept.\  {\bf 317}, 143 (1999)
  [hep-ph/9805494].
  %%CITATION = HEP-PH/9805494;%%
  %361 citations counted in INSPIRE as of 29 Oct 2013
  
  %\cite{ATLAS:2013jma}
\bibitem{ATLAS:2013jma} 
  [ATLAS Collaboration],
  %``Search for high-mass dilepton resonances in 20~$fb^{-1}$ of $pp$ collisions at $\sqrt s = 8$~TeV with the ATLAS experiment,''
  ATLAS-CONF-2013-017.
  %%CITATION = ATLAS-CONF-2013-017;%%
  %29 citations counted in INSPIRE as of 29 Oct 2013
  
  %\cite{CMS:2013qca}
\bibitem{CMS:2013qca} 
  CMS Collaboration [CMS Collaboration],
  %``Search for Resonances in the Dilepton Mass Distribution in pp Collisions at sqrt(s) = 8 TeV,''
  CMS-PAS-EXO-12-061.
  %%CITATION = CMS-PAS-EXO-12-061;%%
  %14 citations counted in INSPIRE as of 29 Oct 2013
  
  %\cite{Godfrey:2013eta}
\bibitem{Godfrey:2013eta} 
  S.~Godfrey and T.~Martin,
  %``Z' Discovery Reach at Future Hadron Colliders: A Snowmass White Paper,''
  arXiv:1309.1688 [hep-ph].
  %%CITATION = ARXIV:1309.1688;%%
  %4 citations counted in INSPIRE as of 05 May 2014
  
  %\cite{LEP:2003aa}
\bibitem{LEP:2003aa} 
  t.~S.~Electroweak [LEP and ALEPH and DELPHI and L3 and OPAL and LEP Electroweak Working Group and SLD Electroweak Group and SLD Heavy Flavor Group Collaborations],
  %``A Combination of preliminary electroweak measurements and constraints on the standard model,''
  hep-ex/0312023.
  %%CITATION = HEP-EX/0312023;%%
  %218 citations counted in INSPIRE as of 29 Oct 2013

%\cite{Carena:2004xs}
\bibitem{Carena:2004xs} 
  M.~S.~Carena, A.~Daleo, B.~A.~Dobrescu and T.~M.~P.~Tait,
  %``$Z^\prime$ gauge bosons at the Tevatron,''
  Phys.\ Rev.\ D {\bf 70}, 093009 (2004)
  [hep-ph/0408098].
  %%CITATION = HEP-PH/0408098;%%
  %256 citations counted in INSPIRE as of 29 Oct 2013

%%% ILC Studies

  %\cite{Rizzo:1996rx}
\bibitem{Rizzo:1996rx} 
  T.~G.~Rizzo,
  %``An Exploration of below threshold Z-prime mass and coupling determinations at the NLC,''
  Phys.\ Rev.\ D {\bf 55}, 5483 (1997)
  [hep-ph/9612304].
  %%CITATION = HEP-PH/9612304;%%
  %18 citations counted in INSPIRE as of 14 Dec 2013  
  
    %\cite{Rizzo:1998vf}
\bibitem{Rizzo:1998vf} 
  T.~G.~Rizzo,
  %``Distinguishing indirect signatures of new physics at the NLC: Z-prime versus R-parity violation,''
  Phys.\ Rev.\ D {\bf 59}, 113004 (1999)
  [hep-ph/9811440].
  %%CITATION = HEP-PH/9811440;%%
  %31 citations counted in INSPIRE as of 14 Dec 2013
  
  %\cite{Rizzo:2003ug}
\bibitem{Rizzo:2003ug} 
  T.~G.~Rizzo,
  %``Kaluza-Klein / Z-prime differentiation at the LHC and linear collider,''
  JHEP {\bf 0306}, 021 (2003)
  [hep-ph/0305077].
  %%CITATION = HEP-PH/0305077;%%
  %17 citations counted in INSPIRE as of 14 Dec 2013

%\cite{Freitas:2004hq}
\bibitem{Freitas:2004hq} 
  A.~Freitas,
  %``Weakly coupled neutral gauge bosons at future linear colliders,''
  Phys.\ Rev.\ D {\bf 70}, 015008 (2004)
  [hep-ph/0403288].
  %%CITATION = HEP-PH/0403288;%%
  %16 citations counted in INSPIRE as of 29 Oct 2013
  
  \bibitem{Basso:2009hf}
   L.~Basso, A.~Belyaev, S.~Moretti and G.~M.~Pruna,
   %``Probing the Z-prime sector of the minimal B-L model at future Linear Colliders in the e+ e- ---> mu+ mu- process,''
   JHEP {\bf 0910} (2009) 006
   [arXiv:0903.4777 [hep-ph]].
  
  %\cite{Blaising:2012tz}
\bibitem{Blaising:2012tz} 
  J.~-J.~Blaising and J.~D.~Wells,
  %``Physics performances for Z' searches at 3 TeV and 1.5 TeV CLIC,''
  arXiv:1208.1148 [hep-ph].
  %%CITATION = ARXIV:1208.1148;%%
  %2 citations counted in INSPIRE as of 29 Oct 2013
  
  %\cite{Baer:2013cma}
\bibitem{Baer:2013cma} 
  H.~Baer, T.~Barklow, K.~Fujii, Y.~Gao, A.~Hoang, S.~Kanemura, J.~List and H.~E.~Logan {\it et al.},
  %``The International Linear Collider Technical Design Report - Volume 2: Physics,''
  arXiv:1306.6352 [hep-ph].
  %%CITATION = ARXIV:1306.6352;%%
  %28 citations counted in INSPIRE as of 29 Oct 2013
  
%%% Higgs and top

%\cite{Aad:2012tfa}
\bibitem{Aad:2012tfa} 
  G.~Aad {\it et al.}  [ATLAS Collaboration],
  %``Observation of a new particle in the search for the Standard Model Higgs boson with the ATLAS detector at the LHC,''
  Phys.\ Lett.\ B {\bf 716}, 1 (2012)
  [arXiv:1207.7214 [hep-ex]].
  %%CITATION = ARXIV:1207.7214;%%
  %1835 citations counted in INSPIRE as of 29 Oct 2013
  
  %\cite{Chatrchyan:2012ufa}
\bibitem{Chatrchyan:2012ufa} 
  S.~Chatrchyan {\it et al.}  [CMS Collaboration],
  %``Observation of a new boson at a mass of 125 GeV with the CMS experiment at the LHC,''
  Phys.\ Lett.\ B {\bf 716}, 30 (2012)
  [arXiv:1207.7235 [hep-ex]].
  %%CITATION = ARXIV:1207.7235;%%
  %1816 citations counted in INSPIRE as of 29 Oct 2013
  
 %\cite{Peskin:2012we}
\bibitem{Peskin:2012we} 
  M.~E.~Peskin,
  %``Comparison of LHC and ILC Capabilities for Higgs Boson Coupling Measurements,''
  arXiv:1207.2516 [hep-ph].
  %%CITATION = ARXIV:1207.2516;%%
  %64 citations counted in INSPIRE as of 29 Oct 2013 
  
  %\cite{Klute:2013cx}
\bibitem{Klute:2013cx} 
  M.~Klute, R.~Lafaye, T.~Plehn, M.~Rauch and D.~Zerwas,
  %``Measuring Higgs Couplings at a Linear Collider,''
  Europhys.\ Lett.\  {\bf 101}, 51001 (2013)
  [arXiv:1301.1322 [hep-ph]].
  %%CITATION = ARXIV:1301.1322;%%
  %18 citations counted in INSPIRE as of 29 Oct 2013
  
%\cite{Batra:2006iq}
\bibitem{Batra:2006iq} 
  P.~Batra and T.~M.~P.~Tait,
  %``Measuring the W-t-b Interaction at the ILC,''
  Phys.\ Rev.\ D {\bf 74}, 054021 (2006)
  [hep-ph/0606068].
  %%CITATION = HEP-PH/0606068;%%
  %11 citations counted in INSPIRE as of 29 Oct 2013  
  
    %\cite{Alwall:2011uj}
\bibitem{Alwall:2011uj} 
  J.~Alwall, M.~Herquet, F.~Maltoni, O.~Mattelaer and T.~Stelzer,
  %``MadGraph 5 : Going Beyond,''
  JHEP {\bf 1106}, 128 (2011)
  [arXiv:1106.0522 [hep-ph]].
  %%CITATION = ARXIV:1106.0522;%%
  %912 citations counted in INSPIRE as of 29 Oct 2013
  
  %\cite{Sjostrand:2007gs}
\bibitem{Sjostrand:2007gs} 
  T.~Sjostrand, S.~Mrenna and P.~Z.~Skands,
  %``A Brief Introduction to PYTHIA 8.1,''
  Comput.\ Phys.\ Commun.\  {\bf 178}, 852 (2008)
  [arXiv:0710.3820 [hep-ph]].
  %%CITATION = ARXIV:0710.3820;%%
  %911 citations counted in INSPIRE as of 29 Oct 2013
  
    \bibitem{PGS}
http://www.physics.ucdavis.edu/$\sim$conway/research/ software/pgs/pgs4-general.htm

%%% LHC Section

%\cite{Yu:2013wta}
\bibitem{Yu:2013wta} 
  F.~Yu,
  %``Di-jet resonances at future hadron colliders: A Snowmass whitepaper,''
  arXiv:1308.1077 [hep-ph].
  %%CITATION = ARXIV:1308.1077;%%
  %4 citations counted in INSPIRE as of 06 May 2014
  
%\cite{Barger:2006hm}
\bibitem{Barger:2006hm} 
  V.~Barger, T.~Han and D.~G.~E.~Walker,
  %``Top Quark Pairs at High Invariant Mass: A Model-Independent Discriminator of New Physics at the LHC,''
  Phys.\ Rev.\ Lett.\  {\bf 100}, 031801 (2008)
  [hep-ph/0612016].
  %%CITATION = HEP-PH/0612016;%%
  %75 citations counted in INSPIRE as of 06 May 2014  
  
  %\cite{Agashe:2007ki}
\bibitem{Agashe:2007ki} 
  K.~Agashe, H.~Davoudiasl, S.~Gopalakrishna, T.~Han, G.~-Y.~Huang, G.~Perez, Z.~-G.~Si and A.~Soni,
  %``LHC Signals for Warped Electroweak Neutral Gauge Bosons,''
  Phys.\ Rev.\ D {\bf 76}, 115015 (2007)
  [arXiv:0709.0007 [hep-ph]].
  %%CITATION = ARXIV:0709.0007;%%
  %144 citations counted in INSPIRE as of 06 May 2014
  
%%% Other ILC  

%\cite{Osland:2009dp}
\bibitem{Osland:2009dp} 
  P.~Osland, A.~A.~Pankov and A.~V.~Tsytrinov,
  %``Identification of extra neutral gauge bosons at the International Linear Collider,''
  Eur.\ Phys.\ J.\ C {\bf 67}, 191 (2010)
  [arXiv:0912.2806 [hep-ph]].
  %%CITATION = ARXIV:0912.2806;%%
  %6 citations counted in INSPIRE as of 30 Oct 2013

%\cite{Petriello:2008zr}
\bibitem{Petriello:2008zr} 
  F.~Petriello and S.~Quackenbush,
  %``Measuring $Z^\prime$ couplings at the CERN LHC,''
  Phys.\ Rev.\ D {\bf 77}, 115004 (2008)
  [arXiv:0801.4389 [hep-ph]].
  %%CITATION = ARXIV:0801.4389;%%
  %76 citations counted in INSPIRE as of 29 Oct 2013
  
  %\cite{Li:2009xh}
\bibitem{Li:2009xh} 
  Y.~Li, F.~Petriello and S.~Quackenbush,
  %``Reconstructing a Z-prime Lagrangian using the LHC and low-energy data,''
  Phys.\ Rev.\ D {\bf 80}, 055018 (2009)
  [arXiv:0906.4132 [hep-ph]].
  %%CITATION = ARXIV:0906.4132;%%
  %27 citations counted in INSPIRE as of 29 Oct 2013
  
  %%% Still to link

\end{thebibliography}
\end{document}